\documentclass[aps,prl,twocolumn,showpacs,amsmath,amssymb,floatfix]{revtex4-1}
\pdfoutput=1 
\usepackage{graphicx,color,framed}
\usepackage{hyperref}
\usepackage{times}
\usepackage{enumerate}
\usepackage{lipsum}
\usepackage{slashed}
\usepackage{array}
\usepackage{textcomp}
\usepackage{amsthm}

\hypersetup{
    colorlinks=true, 
    linktoc=all,     
    linkcolor=blue,  
}

\def \beq {\begin{equation}}
\def \eeq {\end{equation}}
\def \beqa {\begin{eqnarray}}
\def \eeqa {\end{eqnarray}}
\def \bseq {\begin{subequations}}
\def \eseq {\end{subequations}}
\newcommand \dg {\dagger}
\newcommand \up {\uparrow}
\newcommand \down {\downarrow}
\newcommand \al {\alpha}

\newcommand \ran {\rangle}
\newcommand \lan {\langle}

\newcommand \ep {\epsilon}

\newcommand \mb {\mathbf}

\newcommand \nnb {\nonumber}
\newcommand \ov {\overline}
\newcommand \td {\tilde}

\begin{document}

\title{Momentum occupation number bounds for interacting fermions}
\author{Matthew F. Lapa}
\email[email address: ]{mlapa@uchicago.edu}
\affiliation{Kadanoff Center for Theoretical Physics, University of Chicago, Chicago, Illinois 60637, USA}

\begin{abstract}

We derive rigorous bounds on the average momentum occupation numbers $\lan n_{\mb{k}\sigma}\ran$ in the 
Hubbard and Kondo models in the ground state and at non-zero temperature ($T>0$) in the grand canonical
ensemble. 
For the Hubbard model with $T>0$ our bound proves that, when interaction strength $\ll k_B T\ll$ Fermi energy,  
$\lan n_{\mb{k}\sigma}\ran$ is guaranteed to be close to its value in a low temperature 
free fermion system. For the Kondo model with any $T>0$ our bound proves that
$\lan n_{\mb{k}\sigma}\ran$ tends to its non-interacting value in the infinite volume limit. In the
ground state case our bounds instead show that $\lan n_{\mb{k}\sigma}\ran$
approaches its non-interacting value as $\mb{k}$ moves away from a certain surface
in momentum space. For the Hubbard model at half-filling on a bipartite lattice, 
this surface coincides with the non-interacting Fermi surface. 
In the Supplemental Material we extend our results to some generalized versions of the
Hubbard and Kondo models. Our proofs use the Fermi statistics of the particles in a fundamental way.

\end{abstract}

\pacs{}

\maketitle

\textit{Introduction:} Under certain conditions, gapless free fermion systems are expected to 
be stable to interactions, in the sense that certain properties of the interacting system resemble those
of the free system. This expectation is the basis for Landau's Fermi liquid 
theory~\cite{AGD}, and it can be justified in some cases using a Renormalization Group 
approach~\cite{benfatto-gallavotti,shankar-RG,polchinski}. 
Perhaps the most famous result along these lines is Luttinger's theorem (LT), which states that
the volume enclosed by the Fermi surface in an interacting system~\footnote{Luttinger used
a particular definition of the interacting Fermi surface that makes sense in the context of
perturbation theory.} is equal to the volume enclosed by the Fermi 
surface in the corresponding free system~\cite{LW2,luttinger,D}.  
While this result is interesting,
most derivations of it rely on unproven assumptions, and so it cannot be expected to hold in 
generic models of interacting fermions.

Luttinger's original work relied on perturbation theory, and so his result may not hold
if perturbation theory is not absolutely convergent~\footnote{In Ref.~\onlinecite{luttinger} Luttinger was 
careful to note that his result depends on the reliability of perturbation theory. It may not hold
if perturbation theory is only an asymptotic series.}. Most recent works on LT 
take a different approach but make other assumptions,  
for example that the system is a Fermi liquid at low energies~\cite{oshikawa,paramekanti-vishwanath, 
praz2005,shastry,ETS,heath2020,wen2021}. There are a few
rigorous results on LT in one-dimensional (1D) systems~\cite{YOA} and in some 2D systems that lack inversion 
symmetry~\cite{FKT1}\footnote{Ref.~\onlinecite{FKT1} is the first in a series of 
papers that proves this result.}. However,
there are also several counterexamples to the 
original statement of LT~\cite{PhysRevLett.81.2966,rosch,phillips-1,phillips-2,
PhysRevLett.114.156402}.

It is useful to think of LT as a stability result that states that the momentum space picture in an
interacting system resembles the picture in the corresponding free system. In this work
we prove stability results along these lines for the Hubbard
and Kondo models in their ground state and at non-zero temperature. Specifically, we derive 
rigorous bounds on the deviation of the average momentum occupation numbers $\lan n_{\mb{k}\sigma}\ran$ from 
their non-interacting values. Our focus on $\lan n_{\mb{k}\sigma}\ran$ is inspired by 
Luttinger's original work~\cite{luttinger}, where he showed (again, using perturbation theory) that 
$\lan n_{\mb{k}\sigma}\ran$ has a discontinuity at the location of the interacting Fermi surface. 
This discontinuity was also rigorously proven to exist in some 2D models without inversion 
symmetry~\cite{FKT1}.

Our rigorous bounds on the $\lan n_{\mb{k}\sigma}\ran$ allow us to prove the following results. 
For the Hubbard model at non-zero temperature our bound proves the existence of a parameter regime of the form 
($\ep_F=$ Fermi energy)
\beq
	\text{interaction strength } \ll k_B T \ll \ep_F \nnb 
\eeq
in which $\lan n_{\mb{k}\sigma}\ran$ is guaranteed to be close to its value in a \emph{low}
temperature free fermion system. For the Kondo model at \emph{any} non-zero temperature our bound proves that
$\lan n_{\mb{k}\sigma}\ran$ tends to its non-interacting value in the infinite volume limit. 
In the ground state case our results show that $\lan n_{\mb{k}\sigma}\ran$
approaches its non-interacting value of $0$ or $1$ as $\mb{k}$ moves away from a certain surface
in momentum space. For the Hubbard model at half-filling on a bipartite lattice,
this surface coincides with the non-interacting Fermi surface. In the ground state and $T>0$ cases 
our results for the Kondo model are much stronger, and this is because the interaction
in the Kondo model only involves a single lattice site.  
In the Supplemental Material (SM) we extend these results to generalized versions of the
Hubbard and Kondo models.

For the Hubbard model our stability results are strongest in the case with $T>0$, and it is useful to
discuss the reason for this. The key physical idea involved is that the system is most likely to 
``look'' like a low temperature free fermion system when 
$k_B T \ll \ep_F$ but $T$ is still above the transition temperature for any low temperature 
instabilities (e.g., a superconducting transition or the Kohn-Luttinger 
instability~\cite{kohn-luttinger,shankar-luttinger}). This idea was
strongly emphasized in a fascinating series of works in the mathematical
physics literature that established stability~\footnote{The stability results in these works show
that the imaginary time Green's functions of the models are \emph{analytic} functions of the interaction 
strength.} of free fermions to interactions in 2D systems at low but non-zero 
temperatures~\cite{salmhofer,DR-PRL,DR1,DR2,BGM1,BGM2,rivasseau-hubbard-1,rivasseau-hubbard-2,
rivasseau-hubbard-3}. Our results on the Hubbard model demonstrate the power of this idea in yet 
another concrete setting.

\textit{Hubbard and Kondo models:} We consider Hubbard and Kondo models on a Bravais lattice $\Lambda$. 
Both models feature spinful fermions, and we denote 
by $c_{\mb{x}\sigma}$ and $c^{\dg}_{\mb{x}\sigma}$ the annihilation and creation operators for 
a fermion of spin $\sigma\in\{\up,\down\}$ on a site $\mb{x}\in\Lambda$. These operators obey the standard 
anticommutation relations $\{c_{\mb{x}\sigma},c_{\mb{y}\tau}\} = 0$ and
$\{c_{\mb{x}\sigma},c^{\dg}_{\mb{y}\tau}\} = \delta_{\mb{x}\mb{y}}\delta_{\sigma\tau}$. 
We also define the Fourier-transformed fermions $c_{\mb{k}\sigma}$ by
$c_{\mb{k}\sigma} = |\Lambda|^{-\frac{1}{2}}\sum_{\mb{x}} c_{\mb{x}\sigma}e^{-i \mb{k}\cdot\mb{x}}$,
where $\mb{k}$ is a wave vector in the first Brillouin zone of $\Lambda$, and $|\Lambda|$ is the total number
of sites in the lattice. We define the number operators in real space and reciprocal
space by $n_{\mb{x}\sigma} = c^{\dg}_{\mb{x}\sigma}c_{\mb{x}\sigma}$ and 
$n_{\mb{k}\sigma} = c^{\dg}_{\mb{k}\sigma}c_{\mb{k}\sigma}$. The total number operator is
$\mathcal{N} = \sum_{\sigma}\mathcal{N}_{\sigma}$, where 
$\mathcal{N}_{\sigma} = \sum_{\mb{x}}n_{\mb{x}\sigma} = \sum_{\mb{k}}n_{\mb{k}\sigma}$ is the
number operator for spin $\sigma$. Finally, the Kondo model features an additional
impurity spin of magnitude $s$, with $s\in\{1/2,1,3/2,\dots\}$. 
This spin is represented by the vector operator
$\vec{S}= (S^x, S^y, S^z)$ whose components satisfy the usual relations $[S^x,S^y] = i S^z$ (plus
cyclic permutations) and $\vec{S}\cdot\vec{S} = s(s+1)$.

The Hamiltonians for our models take the form
\begin{subequations}
\label{eq:models}
\begin{align}
	H_{\text{Hubbard}} &= \sum_{\mb{k},\sigma}(\ep_{\mb{k}}-\mu)n_{\mb{k}\sigma} + u \sum_{\mb{x}}n_{\mb{x}\up}n_{\mb{x}\down} \label{eq:Hubbard} \\
	H_{\text{Kondo}} &= \sum_{\mb{k},\sigma}(\ep_{\mb{k}}-\mu)n_{\mb{k}\sigma} + J \vec{S}\cdot \left(\sum_{\tau,\tau'} c^{\dg}_{\mb{0}\tau}\frac{\vec{\sigma}_{\tau\tau'}}{2}c_{\mb{0}\tau'}\right)\ , \label{eq:Kondo}
\end{align}
\end{subequations}
where the different quantities appearing here are as follows. First, $\mu$ is the
chemical potential. Next, the energy dispersion 
$\ep_{\mb{k}}$ is the Fourier transform of (the negative of) a translation invariant hopping matrix
$t_{\mb{x},\mb{y}}$. We assume that $t_{\mb{x},\mb{y}}$ satisfies 
$t_{\mb{x},\mb{x}} = 0$ and $t_{\mb{x},\mb{y}} = t^*_{\mb{y},\mb{x}} = t_{\mb{x} + \mb{r},\mb{y} + \mb{r}}$ 
for any $\mb{r} \in \Lambda$, and then 
$\ep_{\mb{k}} := -\sum_{\mb{r}}t_{\mb{x} + \mb{r},\mb{x}}e^{-i\mb{k}\cdot\mb{r}}$ (which is independent of 
$\mb{x}$ by translation invariance). 
For example, with nearest-neighbor hopping of strength $t/2$ on the (hyper)cubic lattice in $D$ dimensions, 
we have $\ep_{\mb{k}} = -t\sum_{j=1}^D\cos(k_j)$, where the $k_j$ are the components of $\mb{k}$.
Next, $H_{\text{Hubbard}}$ features an on-site Hubbard 
interaction of strength $u$, where $u>0$ ($u<0$) for repulsive (attractive) interactions.
Finally, $H_{\text{Kondo}}$ features a Heisenberg interaction between the impurity spin
and the spin of the fermion at $\mb{x}=\mb{0}$ (here, 
$\vec{\sigma} = (\sigma^x,\sigma^y,\sigma^z)$ is the vector of Pauli matrices, 
$\sigma^x_{\tau\tau'}$ is the $\tau,\tau'$ matrix element
of $\sigma^x$, etc.). This interaction has a strength $J$ and is antiferromagnetic (ferromagnetic)
for $J>0$ ($J<0$).

Our main results concern the expectation values $\lan n_{\mb{k}\sigma}\ran$ in the ground state
of these models and in the thermal state at inverse temperature $\beta = (k_B T)^{-1}$. In both
cases we work in the \emph{grand canonical ensemble}. In the ground state case this means that we
work with the lowest energy state of the Hamiltonian over all possible fermion number sectors. In the
non-zero temperature case this means that we trace over the entire Fock space of the spin-up and spin-down
fermions (in the Kondo case we also trace over the Hilbert space of the impurity spin).

We use $|\psi\ran$ to denote the grand canonical ground state or a particular ground state if there is
a ground state degeneracy. The expectation value of any operator
$\mathcal{O}$ is defined by $\lan \mathcal{O}\ran := \lan\psi|\mathcal{O}|\psi\ran$ in the ground state
case and at non-zero temperature by 
$\lan \mathcal{O}\ran := \text{Tr}\left[ \mathcal{O} e^{-\beta H}\right]/ Z$, where 
$Z= \text{Tr}\left[e^{-\beta H}\right]$.
For the Hubbard model the ground state can always be chosen to have a 
definite number of fermions of each spin~\footnote{If the model has a ground state degeneracy, then
we can always choose a basis for the ground state subspace such that each state is a simultaneous eigenstate
of $\mathcal{N}_{\up}$ and $\mathcal{N}_{\down}$.}. We denote these fermion numbers by $N_{\sigma}$ and the
corresponding filling fractions by $\rho_{\sigma}$, i.e., 
$\mathcal{N}_{\sigma}|\psi\ran = N_{\sigma}|\psi\ran$ and $\rho_{\sigma} = N_{\sigma}/|\Lambda|$. With 
this notation, we are now ready to present our results.

\textit{Theorem 1 (non-zero temperature):} Let $f_{\mb{k}}$ denote the Fermi-Dirac 
distribution with chemical potential $\mu$, 
$f_{\mb{k}} = (e^{\beta (\ep_{\mb{k}} - \mu)} + 1)^{-1}$.
For any $\beta < \infty$, the momentum occupation numbers $\lan n_{\mb{k}\sigma}\ran$ for the models
in Eq.~\eqref{eq:models} obey
\beq
	-\delta f_{\mb{k}} \leq \lan n_{\mb{k}\sigma}\ran - f_{\mb{k}} \leq \delta(1 - f_{\mb{k}})\ ,
\eeq
where the constant $\delta$ is given by
\beq
	\delta = \begin{cases}
		\beta |u| & ,\text{ Hubbard model} \\
		\beta \frac{3|J|s}{2\sqrt{|\Lambda|}} & ,\text{ Kondo model}\ .
	\end{cases}
\eeq

\textit{Discussion:} The momentum occupation
numbers for the free model $H_0 = \sum_{\mb{k},\sigma}(\ep_{\mb{k}} - \mu)n_{\mb{k}\sigma}$
are given exactly by the Fermi-Dirac distribution $f_{\mb{k}}$. Therefore, 
Theorem 1 shows that, when $\delta \ll 1$, the momentum occupation numbers
for the interacting system are very close to those of the
free model $H_0$. In the Kondo case we also have $\delta\to 0$ in the infinite volume limit
$|\Lambda|\to\infty$, 
so for that model $\lan n_{\mb{k}\sigma}\ran \to f_{\mb{k}}$ as $|\Lambda|\to\infty$ at any
non-zero temperature.

Let us now consider the Hubbard model. In that case $\delta \ll 1$ will hold at high 
temperatures, but the most interesting aspect of Theorem 1
is that it reveals the existence of a regime where the $\lan n_{\mb{k}\sigma}\ran$ resemble the 
occupation numbers of a free fermion system at \emph{low} temperature. To see this,
recall that the free system described by $H_0$ is said to be at low temperature if 
$k_B T \ll \ep_{\text{F}}$, where 
$\ep_{\text{F}} := |\ep_{\mb{0}} - \mu|$ is the Fermi energy
and $\ep_{\mb{0}}$ is the value of the dispersion at the origin 
($\mb{k} = \mb{0}$) of the Brillouin zone. Then Theorem 1 implies that the $\lan n_{\mb{k}\sigma}\ran$ 
resemble the momentum occupation numbers of a low temperature free fermion system if
$u,T$, and $\ep_F$ obey $|u| \ll k_B T \ll \ep_F$.

\textit{Theorem 2 (Hubbard, ground state):} In any ground state of $H_{\text{Hubbard}}$ 
the momentum occupation numbers $\lan n_{\mb{k}\sigma}\ran$ obey 
\begin{subequations}
\begin{align}
	\lan n_{\mb{k}\sigma}\ran &\leq \frac{|u|\sqrt{\rho_{\ov{\sigma}}}}{\ep_{\mb{k}} - \mu + u\rho_{\ov{\sigma}}}\ \ ,\ \text{ if }\ \ep_{\mb{k}} - \mu + u\rho_{\ov{\sigma}} > 0 \\
	1 - \lan n_{\mb{k}\sigma}\ran &\leq \frac{|u|\sqrt{\rho_{\ov{\sigma}}}}{|\ep_{\mb{k}} - \mu + u\rho_{\ov{\sigma}}|}\ \ ,\ \text{ if }\ \ep_{\mb{k}} - \mu + u\rho_{\ov{\sigma}} < 0\ ,
\end{align}
\end{subequations}
where $\ov{\sigma}$ is the opposite of $\sigma$ (e.g., $\ov{\up} = \down$).

\textit{Discussion:} These bounds show that $\lan n_{\mb{k}\sigma}\ran$ approaches its non-interacting 
value of $0$ or $1$ as $\mb{k}$ moves away from the surface in reciprocal space defined by
$\ep_{\mb{k}} - \mu + u\rho_{\ov{\sigma}} = 0$ (note that the non-interacting Fermi surface
is defined by $\ep_{\mb{k}} - \mu = 0$). 
There is also a small region around this surface where these bounds are no longer effective 
because the denominator becomes smaller
than the numerator as $\mb{k}$ approaches
this surface. The size of this region is determined by the interaction strength 
$u$ and the densities $\rho_{\sigma}$. Finally, this bound has an interesting property in the
case of half-filling on a bipartite lattice~\footnote{In this case
we assume that the $t_{\mb{x},\mb{y}}$ are real and that $t_{\mb{x},\mb{y}} = 0$ if $\mb{x}$ and $\mb{y}$ are 
in the same sublattice of the bipartite lattice~\cite{lieb,LN}.}, where $\mu = u/2$ and 
$\rho_{\sigma} = 1/2$~\cite{lieb,LN}. 
In this case $-\mu + u\rho_{\ov{\sigma}} = 0$ and so the surface defined by 
$\ep_{\mb{k}} - \mu + u\rho_{\ov{\sigma}} = 0$
coincides with the non-interacting Fermi surface at half-filling, which is just defined by
$\ep_{\mb{k}}= 0$.

\textit{Theorem 3 (Kondo, ground state):} In any ground state of $H_{\text{Kondo}}$ 
the momentum occupation numbers $\lan n_{\mb{k}\sigma}\ran$ obey 
\begin{subequations}
\begin{align}
	\lan n_{\mb{k}\sigma}\ran &\leq \frac{3}{2\sqrt{|\Lambda|}}\frac{|J|s}{\ep_{\mb{k}} - \mu}\ \ ,\ \text{ if }\ \ep_{\mb{k}} - \mu  > 0 \\
	1 - \lan n_{\mb{k}\sigma}\ran &\leq \frac{3}{2\sqrt{|\Lambda|}}\frac{|J|s}{|\ep_{\mb{k}} - \mu|}\ \ ,\ \text{ if }\ \ep_{\mb{k}} - \mu < 0\ .
\end{align}
\end{subequations}

\textit{Discussion:} 
A related result was obtained in Theorem 2 of Ref.~\onlinecite{bravyi-gosset} for a different family of 
quantum impurity models. In comparing with our Theorem 2, this bound has an extra
factor of $\sqrt{|\Lambda|}$, and so it is much more powerful
than our result for the Hubbard model. In particular, for any $\mb{k}$ that is far enough from the
non-interacting Fermi surface to satisfy an inequality of the form 
\beq
	|\ep_{\mb{k}} - \mu| \geq A |\Lambda|^{-p}\ \ ,\ \ p < 1/2\ , \label{eq:distance-bound}
\eeq
where $A$ is a constant with units of energy, we find that $\lan n_{\mb{k}\sigma}\ran$ tends to its 
non-interacting value of $0$ or $1$ in the 
infinite volume limit. The only values of $\mb{k}$ that do not satisfy a bound like
\eqref{eq:distance-bound} are contained within a small region around the non-interacting Fermi surface, and
the width of this region vanishes in the limit $|\Lambda|\to\infty$.

\textit{Plan for the rest of the main text:} In the rest of the main text we present the proof
of Theorem 1 for the Hubbard model. We prove our other results in the SM.
The key to proving Theorem 1 is a basic bound that we state in Lemma 1 below. 
We now state Lemma 1 and then use it to prove Theorem 1. We then present the proof of Lemma 1 itself.

\textit{Lemma 1:} 
Let $|\phi\ran$ be any normalized state in the Fock space of the spin-up and spin-down
fermions, and let $U = u \sum_{\mb{x}}n_{\mb{x}\up}n_{\mb{x}\down}$ be the Hubbard
interaction. Then for any $\mb{k}$ and $\sigma$ the expectation value
$\lan \phi| c^{\dg}_{\mb{k}\sigma}	[U,c_{\mb{k}\sigma}]|\phi\ran$ obeys 
\beq
	|\lan \phi| c^{\dg}_{\mb{k}\sigma}	[U,c_{\mb{k}\sigma}]|\phi\ran| \leq |u|\ , \label{eq:lemma-1-bound}
\eeq
and an identical bound holds for $|\lan \phi| c_{\mb{k}\sigma}	[U,c^{\dg}_{\mb{k}\sigma}]|\phi\ran|$.

\textit{Remark:} The same bound holds for the thermal expectation value
$\lan c^{\dg}_{\mb{k}\sigma}	[U,c_{\mb{k}\sigma}]\ran$. To see it, let $|\ell\ran$ and $E_{\ell}$ be a 
complete set of eigenvectors and eigenvalues of $H_{\text{Hubbard}}$. Then we have
\begin{align}
	|\lan c^{\dg}_{\mb{k}\sigma}	[U,c_{\mb{k}\sigma}]\ran| &= \Big|\frac{1}{Z}\sum_{\ell}\lan \ell| c^{\dg}_{\mb{k}\sigma}	[U,c_{\mb{k}\sigma}]|\ell\ran e^{-\beta E_{\ell}}\Big| \nnb \\
 &\leq \frac{1}{Z}\sum_{\ell}|\lan \ell| c^{\dg}_{\mb{k}\sigma}	[U,c_{\mb{k}\sigma}]|\ell\ran| e^{-\beta E_{\ell}} \nnb \\
 &\leq |u|\ ,
\end{align}
where the last line follows from $Z = \sum_{\ell}e^{-\beta E_{\ell}}$. 

\textit{Proof of Theorem 1 (Hubbard case):} The first step is to use the thermodynamic inequality
\begin{align}
	\frac{1}{2}\beta\lan \mathcal{O}^{\dg}[H,\mathcal{O}] - [H, \mathcal{O}^{\dg}]\mathcal{O}\ran \geq \Phi(\lan\mathcal{O}^{\dg}\mathcal{O}\ran, \lan\mathcal{O}\mathcal{O}^{\dg}\ran)\ ,
\end{align}
where $\mathcal{O}$ can be any operator and
$\Phi(u,v)$ is the function of two real variables defined by
\beq
	\Phi(u,v) := u\ln(u)-u\ln(v)\ .
\eeq
This inequality is a local version of the \emph{Gibbs variational principle}, 
and it can be derived as in Lemma 6 of Ref.~\onlinecite{sewell}. 

We use this inequality twice: first with $\mathcal{O} = c_{\mb{k}\sigma}$, and
then with $\mathcal{O} = c^{\dg}_{\mb{k}\sigma}$. In the first case we find that
\begin{align}
	-\beta(\ep_{\mb{k}}-\mu)\lan n_{\mb{k}\sigma} \ran &+ \frac{1}{2}\beta\lan c^{\dg}_{\mb{k}\sigma}
	[U,c_{\mb{k}\sigma}] - [U, c_{\mb{k}\sigma}^{\dg}]c_{\mb{k}\sigma}\ran \nnb \\
	&\geq \Phi(\lan n_{\mb{k}\sigma} \ran, 1-\lan n_{\mb{k}\sigma} \ran) \label{eq:first-ineq}
\end{align}
and in the second case we find that
\begin{align}
	\beta(\ep_{\mb{k}}-\mu)(1-\lan n_{\mb{k}\sigma} \ran) &+ \frac{1}{2}\beta\lan c_{\mb{k}\sigma}
	[U,c^{\dg}_{\mb{k}\sigma}] - [U, c_{\mb{k}\sigma}]c^{\dg}_{\mb{k}\sigma}\ran \nnb \\
	&\geq \Phi(1-\lan n_{\mb{k}\sigma} \ran, \lan n_{\mb{k}\sigma} \ran)\ .
\end{align}

Next, we use Lemma 1 to obtain upper bounds on the terms involving $U$ in these inequalities. 
For example, in \eqref{eq:first-ineq} we can use
\begin{align}
	\frac{1}{2}\lan c^{\dg}_{\mb{k}\sigma}
	[U,c_{\mb{k}\sigma}] - [U, c_{\mb{k}\sigma}^{\dg}]c_{\mb{k}\sigma}\ran \leq |\lan c^{\dg}_{\mb{k}\sigma}	[U,c_{\mb{k}\sigma}]\ran| \leq |u|\ .
\end{align}
After applying Lemma 1 our two inequalities take the form
\begin{subequations}
\label{eq:two-inequalities}
\begin{align}
	-\beta(\ep_{\mb{k}}-\mu)\lan n_{\mb{k}\sigma} \ran + \delta &\geq \Phi(\lan n_{\mb{k}\sigma} \ran, 1-\lan n_{\mb{k}\sigma} \ran) \\
	\beta(\ep_{\mb{k}}-\mu)(1-\lan n_{\mb{k}\sigma} \ran) + \delta 
	&\geq \Phi(1-\lan n_{\mb{k}\sigma} \ran, \lan n_{\mb{k}\sigma} \ran)\ ,
\end{align}
\end{subequations}
where $\delta = \beta|u|$ as before.

To complete the proof we need to use inequalities \eqref{eq:two-inequalities}
to obtain upper and lower bounds on the difference $\lan n_{\mb{k}\sigma}\ran - f_{\mb{k}}$. 
To do this we use the fact that $\Phi(x,1-x)$ and $\Phi(1-x,x)$ are 
both convex functions of $x$ for $x\in(0,1)$. A convex function $f(x)$ obeys the bound
$f(x) \geq f(x_0) + f'(x_0)(x-x_0)$ for any $x_0\neq x$, where $f'(x) = df(x)/dx$. We now apply this  
bound to the inequalities in Eq.~\eqref{eq:two-inequalities}, taking $x = \lan n_{\mb{k}\sigma}\ran$ and 
$x_0 = f_{\mb{k}}$. For the first inequality in Eq.~\eqref{eq:two-inequalities} we have
$f(x) = \Phi(x, 1-x)$, $f(x_0) = -\beta(\ep_{\mb{k}}-\mu)x_0$, and 
$f'(x_0) = 1 - \beta(\ep_{\mb{k}}-\mu) + e^{-\beta(\ep_{\mb{k}}-\mu)}$. 
We then find, after some algebra, that 
\beq
	\lan n_{\mb{k}\sigma}\ran - f_{\mb{k}} \leq \delta(1-f_{\mb{k}})\ .
\eeq 
For the second inequality in Eq.~\eqref{eq:two-inequalities} we instead have $f(x) = \Phi(1-x,x)$, and
in that case we find that
\beq
	\lan n_{\mb{k}\sigma}\ran - f_{\mb{k}} \geq -\delta f_{\mb{k}}\ .
\eeq
These two inequalities complete the proof of Theorem 1.

\textit{Proof of Lemma 1:} We will prove the bound in 
Eq.~\eqref{eq:lemma-1-bound} for the case of spin-up. The proofs for the other bounds in Lemma 1 are
nearly identical.
We start by defining new fermion operators $\td{c}_{\mb{x}\sigma}$ that
obey all the usual anticommutation relations except that $\td{c}_{\mb{x}\up}$ now \emph{commutes} with
$\td{c}_{\mb{y}\down}$ and $\td{c}^{\dg}_{\mb{y}\down}$ for all $\mb{x}$ and $\mb{y}$. These operators are 
defined as $\td{c}_{\mb{x}\up} = c_{\mb{x}\up}$ and 
$\td{c}_{\mb{x}\down} = (-1)^{\mathcal{N}_{\up}} c_{\mb{x}\down}$. In
terms of these we also define the number operators 
$\td{n}_{\mb{x}\sigma} = \td{c}^{\dg}_{\mb{x}\sigma}\td{c}_{\mb{x}\sigma}$, which are equal to 
the original operators $n_{\mb{x}\sigma}$. We also
define Fourier-transformed operators $\td{c}_{\mb{k}\sigma}$ and their number
operators $\td{n}_{\mb{k}\sigma} = \td{c}^{\dg}_{\mb{k}\sigma}\td{c}_{\mb{k}\sigma} = n_{\mb{k}\sigma}$ 
exactly as before. The Hubbard model has the interesting property that it takes the same form 
when expressed in terms of these new operators. 
In addition, we can now view $H_{\text{Hubbard}}$ as acting on the tensor product 
$\td{\mathcal{F}}_{\up}\otimes\td{\mathcal{F}}_{\down}$ 
of the Fock spaces $\td{\mathcal{F}}_{\up}$ and $\td{\mathcal{F}}_{\down}$ for the
new spin-up and spin-down fermions $\td{c}_{\mb{x}\up}$ and $\td{c}_{\mb{x}\down}$. We will use this tensor
product structure shortly.

To bound $|\lan \phi| c^{\dg}_{\mb{k}\up}[U,c_{\mb{k}\up}]|\phi\ran|$, we start with the explicit 
formula
\begin{align}
	c^{\dg}_{\mb{k}\up}[U,c_{\mb{k}\up}] = -\frac{u}{|\Lambda|}\sum_{\mb{x},\mb{q}}\td{c}^{\dg}_{\mb{k}\up}\td{c}_{\mb{q}\up}\td{n}_{\mb{x}\down}e^{-i(\mb{k}-\mb{q})\cdot\mb{x}}\ .
	\label{eq:expectation-formula}
\end{align}
At this point it is useful to explain why it is slightly subtle to obtain a $|\Lambda|$-independent bound on 
$|\lan \phi| c^{\dg}_{\mb{k}\up}[U,c_{\mb{k}\up}]|\phi\ran|$. We can see from 
Eq.~\eqref{eq:expectation-formula} that this quantity
has one factor of $|\Lambda|$ in the denominator, but \emph{two} sums over $|\Lambda|$ terms each 
(the sums over $\mb{x}$ and $\mb{q}$), and the absolute value of the summand
$\lan \phi| \td{c}^{\dg}_{\mb{k}\up}\td{c}_{\mb{q}\up}\td{n}_{\mb{x}\down}|\phi\ran 
e^{-i(\mb{k}-\mb{q})\cdot\mb{x}}$
is (naively) of order $1$. From this simple analysis it seems like we will end
up with a bound on this quantity of order $|\Lambda|$. This analysis is incorrect because 
it does not take into account cancellations that follow from Fermi statistics. 

To proceed, we expand $|\phi\ran$ in a way that uses the tensor product structure 
of the Hilbert space when we work in terms of the new fermion operators $\td{c}_{\mb{x}\sigma}$. 
Let $\{|a\ran\}$ be a real space occupation
number basis for $\td{\mathcal{F}}_{\up}$, 
and let $\{|\al\ran\}$ be the same for $\td{\mathcal{F}}_{\down}$. To be more precise, each
state $|a\ran$ is determined by $|\Lambda|$ different numbers $a_{\mb{x}}\in\{0,1\}$ that satisfy 
$\sum_{\mb{x}}a_{\mb{x}} \leq |\Lambda|$, and $|a\ran$ takes the form
$|a\ran = \prod_{\mb{x}}(\td{c}^{\dg}_{\mb{x}\up})^{a_{\mb{x}}}|0\ran_{\up}$, where $|0\ran_{\up}$ is 
the Fock vacuum for $\td{\mathcal{F}}_{\up}$ and where the order of the product is not important here. 
The states $|\al\ran$ for $\td{\mathcal{F}}_{\down}$ take a similar form.

Using these basis states, we expand $|\phi\ran$ as
\beq
	|\phi\ran = \sum_{a,\al}W_{a\al}|a\ran \otimes |\alpha\ran\ ,
\eeq
where $W_{a\al}$ is a matrix of coefficients (this step is inspired by Ref.~\onlinecite{lieb}). 
Since $|\phi\ran$ is normalized, the coefficients $W_{a\al}$ obey the sum rule $\sum_{a,\al}|W_{a\al}|^2 = 1$. 
Since $\td{n}_{\mb{x}\down}$ is diagonal in the $|\al\ran$ basis, we now find that
\begin{align}
	\lan \phi| c^{\dg}_{\mb{k}\up}&[U,c_{\mb{k}\up}]|\phi\ran = \nnb \\
	 -&\frac{u}{|\Lambda|}\sum_{\mb{x}, \mb{q}} e^{-i(\mb{k}-\mb{q})\cdot\mb{x}} \sum_{\al} \lan\chi_{\al}|\td{c}^{\dg}_{\mb{k}\up}\td{c}_{\mb{q}\up}|\chi_{\al}\ran  \lan\al|\td{n}_{\mb{x}\down}|\al\ran\ ,
\end{align}
where we defined the states $|\chi_{\al}\ran\in\td{\mathcal{F}}_{\up}$ by
\beq
	|\chi_{\al}\ran := \sum_a W_{a\al}|a\ran\ .
\eeq
These states are \emph{not} normalized. Instead, their norms satisfy the sum rule
\beq
	\sum_{\al}\lan\chi_{\al}|\chi_{\al}\ran = 1\ . \label{eq:norms}
\eeq
If we also define the coefficients $A^{(\al)}_{\mb{k}\mb{q}}$ and $M^{(\al)}_{\mb{k}\mb{q}}$ by
\begin{subequations}
\begin{align}
	A^{(\al)}_{\mb{k}\mb{q}} &:= u\sum_{\mb{x}}\lan\al|\td{n}_{\mb{x}\down}|\al\ran e^{-i(\mb{k}-\mb{q})\cdot\mb{x}} \\
	M^{(\al)}_{\mb{k}\mb{q}} &:= \lan\chi_{\al}|\td{c}^{\dg}_{\mb{k}\up}\td{c}_{\mb{q}\up}|\chi_{\al}\ran\ ,
\end{align}
\end{subequations}
then at this point we have
\beq
	\lan\phi| c^{\dg}_{\mb{k}\up}[U,c_{\mb{k}\up}]|\phi\ran = -\frac{1}{|\Lambda|}\sum_{\al}\sum_{\mb{q}}A^{(\al)}_{\mb{k}\mb{q}}M^{(\al)}_{\mb{k}\mb{q}}\ ,
\eeq
where we have exchanged the order of the sums.

We now use the triangle inequality on the outer sum over $\al$ to obtain
\beq
	|\lan\phi| c^{\dg}_{\mb{k}\up}[U,c_{\mb{k}\up}]|\phi\ran| \leq \frac{1}{|\Lambda|}\sum_{\al}\Big|\sum_{\mb{q}}A^{(\al)}_{\mb{k}\mb{q}}M^{(\al)}_{\mb{k}\mb{q}}\Big|\ .
\eeq
We then bound the inner sum over $\mb{q}$ using the Cauchy-Schwarz inequality, 
\begin{align}
	\Big|\sum_{\mb{q}}A^{(\al)}_{\mb{k}\mb{q}}M^{(\al)}_{\mb{k}\mb{q}}\Big| \leq \sqrt{\left(\sum_{\mb{q}}|A^{(\al)}_{\mb{k}\mb{q}}|^2\right)\left(\sum_{\mb{q}}|M^{(\al)}_{\mb{k}\mb{q}}|^2\right)}\ .
\end{align}
Next, we have
\begin{align}
	\sum_{\mb{q}}|A^{(\al)}_{\mb{k}\mb{q}}|^2 = u^2|\Lambda|\sum_{\mb{x}}\lan\al|\td{n}_{\mb{x}\down}|\al\ran^2 \leq (u|\Lambda|)^2\ ,
\end{align}
where the first equality is just the Plancherel theorem. The last step is to bound the 
sum involving $M^{(\al)}_{\mb{k}\mb{q}}$. To do that, we need a few facts about
fermion density matrices.

\textit{Fermion density matrices and operator norms:} Consider a set of fermion creation and annihilation 
operators $c_i$, $c^{\dg}_i$ obeying the usual relations $\{c_i,c_j\}=0$ and 
$\{c_i,c^{\dg}_j\} = \delta_{ij}$, where the indices $i$ and $j$
take values in some finite index set $\mathcal{I}$. For any state $|\chi\ran$ in the Fock space
of these operators, we can define a Hermitian matrix $M$ whose matrix elements are given by
$M_{ij} := \lan\chi| c^{\dg}_i c_j|\chi\ran$. This matrix is the single-particle reduced
density matrix for the fermions in the state $|\chi\ran$. An important result about 
$M$ is that, if $\lambda$ is any eigenvalue of $M$, then $0\leq \lambda\leq \lan\chi|\chi\ran$ (we do not
assume that $|\chi\ran$ is normalized)~\cite{yang-ODLRO}. For a short proof of this result, see the SM.

We now review some facts about operator norms of Hermitian matrices. 
The operator norm $||A||$ of a Hermitian matrix $A$ 
is equal to the maximum of the absolute values of the eigenvalues of $A$. For the
fermion density matrix $M$ from the last paragraph, we then find that $||M||\leq \lan\chi|\chi\ran$. 
Next, if $A_{ij}$ is any matrix element of $A$, we have $|A_{ij}|\leq ||A||$ (this follows from the
Cauchy-Schwarz inequality). Finally, the operator norm is \emph{submultiplicative}, which means that
$||AB||\leq ||A||\ ||B||$ for any two matrices $A$ and $B$. 

\textit{Finishing the proof of Lemma 1:} We now use this information to bound the sum
involving $M^{(\al)}_{\mb{k}\mb{q}}$. First, let $M^{(\al)}$ be the Hermitian matrix with matrix elements 
$M^{(\al)}_{\mb{k}\mb{q}}$, and let $N^{(\al)}$ be the square of this matrix, 
$N^{(\al)} = M^{(\al)}M^{(\al)}$. Then 
$\sum_{\mb{q}}|M^{(\al)}_{\mb{k}\mb{q}}|^2 = N^{(\al)}_{\mb{k}\mb{k}} = |N^{(\al)}_{\mb{k}\mb{k}}|$, and we
have
\begin{align}
	\sum_{\mb{q}}|M^{(\al)}_{\mb{k}\mb{q}}|^2 \leq ||N^{(\al)}|| \leq ||M^{(\al)}||^2 \leq \lan\chi_{\al}|\chi_{\al}\ran^2\ .
\end{align}
Combining all of our results leads to 
\beq
	|\lan \phi|c^{\dg}_{\mb{k}\up}[U,c_{\mb{k}\up}]|\phi\ran| \leq \frac{1}{|\Lambda|}\sum_{\al}\sqrt{(u |\Lambda|)^2 \lan\chi_{\al}|\chi_{\al}\ran^2}\ ,
\eeq
and then the bound $|\lan \phi| c^{\dg}_{\mb{k}\up}[U,c_{\mb{k}\up}]|\phi\ran| \leq |u|$
follows from the normalization condition \eqref{eq:norms} for the states $|\chi_{\al}\ran$.

\textit{Conclusion:} For the Hubbard and Kondo models, in the ground state and at non-zero temperature,
we have derived rigorous bounds on the deviation of the average momentum occupation numbers 
$\lan n_{\mb{k}\sigma}\ran$ from their non-interacting values.
In the future it would be interesting to derive similar results for models with more general interactions,
for example spinless fermions with a density-density interaction of the form
$\sum_{\mb{x},\mb{r}}v_{\mb{r}}n_{\mb{x}}n_{\mb{x}+\mb{r}}$ and where the interaction potential
$v_{\mb{r}}$ satisfies a summability condition such as $\sum_{\mb{r}}|v_{\mb{r}}| \leq O(1)$ (or any
similar condition). 
It would also be interesting to try and derive similar bounds for time-dependent quantities
such as Green's functions.

We acknowledge the support of the Kadanoff Center for Theoretical Physics at the 
University of Chicago. This work was supported by the Simons Collaboration on Ultra-Quantum Matter, which is a 
grant from the Simons Foundation (651440).


%

\end{document}


\title{Supplemental Material for ``Momentum occupation number bounds for interacting fermions''}

\author{Matthew F. Lapa}
\email[email address: ]{mlapa@uchicago.edu}
\affiliation{Kadanoff Center for Theoretical Physics, University of Chicago, Chicago, IL, 60637, USA}

\maketitle

\section{Results for generalized Hubbard models}

\subsection{The models}

In this section we study generalized Hubbard models with Hamiltonians of the form
\beq
	H = \sum_{\mb{k},\sigma}\ep_{\mb{k}}n_{\mb{k}\sigma} - \sum_{\mb{x},\sigma}\mu_{\mb{x}}n_{\mb{x}\sigma} + \sum_{\mb{x}}u_{\mb{x}}n_{\mb{x}\up}n_{\mb{x}\down}\ . \label{eq:gen-Hubbard}
\eeq
The main differences between this Hamiltonian and the one from the main text is that we now have a
spatially-varying single-particle potential $\mu_{\mb{x}}$ and a 
spatially-varying Hubbard interaction $u_{\mb{x}}$.

We now introduce some notation that will help us express our results for this generalized model.
First, we define $\ov{\mu}$ and $s_{\mu}$ to be the mean and
standard deviation, respectively, of the coefficients $\mu_{\mb{x}}$,
$\ov{\mu} = \sum_{\mb{x}}\mu_{\mb{x}}/|\Lambda|$ and 
$s_{\mu}^2 = \sum_{\mb{x}}(\mu_{\mb{x}} - \ov{\mu})^2 /|\Lambda|$. If $\mu_{\mb{x}} = \mu$ for all $\mb{x}$
then $\ov{\mu} = \mu$, $s_{\mu} = 0$, and the single-particle potential term in $H$ reduces to the ordinary
chemical potential term $-\mu\mathcal{N}$. However, when the $\mu_{\mb{x}}$ are not uniform the standard
deviation $s_{\mu}$ serves as a natural measure of the disorder in the single-particle potential. 
[We always consider a single Hamiltonian with fixed values of the $\mu_{\mb{x}}$ and $u_{\mb{x}}$, i.e., we 
\emph{do not} do any disorder averaging.] Next, we define $\ov{u}$ and $s_u$ to be the mean
and standard deviation of the interaction coefficients $u_{\mb{x}}$. Finally, we also define
$u_{\text{rms}}$ to be the root mean square of the Hubbard interaction strength,
$u_{\text{rms}}^2 = \sum_{\mb{x}}u_{\mb{x}}^2 /|\Lambda|$. Note that 
$u_{\text{rms}}^2 = \ov{u}^2 + s_u^2$ and so $u_{\text{rms}}\leq \ov{u} + s_u$ by subadditivity of the square
root.

In what follows we write $H$ in the form
\beq
	H = H_0 + V
\eeq
where now 
\beq
	H_0 = \sum_{\mb{k},\sigma}(\ep_{\mb{k}}-\ov{\mu})n_{\mb{k}\sigma}
\eeq
and
\beq
	V = \sum_{\mb{x},\sigma}(\ov{\mu}-\mu_{\mb{x}})n_{\mb{x}\sigma} + \sum_{\mb{x}}u_{\mb{x}}n_{\mb{x}\up}n_{\mb{x}\down}\ . \label{eq:perturbation}
\eeq
Note that the chemical potential in $H_0$ is now the average $\ov{\mu}$ of the coefficients
$\mu_{\mb{x}}$ for the single-particle potential. Then $H_0$ is again the Hamiltonian for
a translation invariant free fermion system, and we can think of $V$ as a perturbation to $H_0$ that 
introduces both interactions and disorder.

\subsection{Non-zero temperature result}

For our generalized Hubbard models at non-zero temperature, we now wish to compare $\lan n_{\mb{k}\sigma}\ran$
with the Fermi-Dirac distribution $f_{\mb{k}}$ with the average chemical potential $\ov{\mu}$. 
In this case our result reads as follows.

\textit{Theorem 1 (generalized Hubbard, non-zero temperature):} Let $f_{\mb{k}}$ denote the Fermi-Dirac 
distribution with chemical potential $\ov{\mu}$, 
$f_{\mb{k}} = (e^{\beta (\ep_{\mb{k}} - \ov{\mu})} + 1)^{-1}$.
For any $\beta < \infty$, the momentum occupation numbers $\lan n_{\mb{k}\sigma}\ran$ for the
Hamiltonian \eqref{eq:gen-Hubbard} obey the bounds
\beq
	-\delta f_{\mb{k}} \leq \lan n_{\mb{k}\sigma}\ran - f_{\mb{k}} \leq \delta(1 - f_{\mb{k}})\ ,
\eeq
where the constant $\delta$ is given by
\beq
	\delta = \beta(s_{\mu} + u_{\text{rms}})\ .
\eeq

We see that in this more general case the deviation of $\lan n_{\mb{k}\sigma}\ran$ from
$f_{\mb{k}}$ is controlled by $s_{\mu}$, which is a natural measure of the disorder in the
single-particle potential, and by $u_{\text{rms}}$, which is a natural measure of the non-uniform 
Hubbard interaction strength. In this case
we again find that $\lan n_{\mb{k}\sigma}\ran$ is close to the result for a low temperature free
fermion system if $s_{\mu} + u_{\text{rms}} \ll k_B T \ll \ep_F$ (recall that
$\ep_F := |\ep_{\mb{0}}-\ov{\mu}|$ is the Fermi energy for $H_0$). In other words, the
presence of disorder has no qualitative effect on our conclusions.

The proof of this theorem follows the same steps as the proof of Theorem 1 in the main text except that
now $H_0$ contains the average potential $\ov{\mu}$, and we also need to replace the
translation invariant Hubbard interaction $U = u\sum_{\mb{x}}n_{\mb{x}\up}n_{\mb{x}\down}$ with
the more general perturbation $V$ from Eq.~\eqref{eq:perturbation}. 
We also need a new version of Lemma 1 for this general case.

\textit{Lemma 1 (generalized Hubbard model):} 
Let $|\phi\ran$ be any normalized state in the Fock space of the spin-up and spin-down
fermions, and let $V$ be the perturbation term from
\eqref{eq:perturbation}. Then for any $\mb{k}$ and $\sigma$ the expectation value
$\lan \phi| c^{\dg}_{\mb{k}\sigma}	[V,c_{\mb{k}\sigma}]|\phi\ran$ obeys 
\beq
	|\lan \phi| c^{\dg}_{\mb{k}\sigma}	[V,c_{\mb{k}\sigma}]|\phi\ran| \leq s_{\mu} + u_{\text{rms}}\ ,
\eeq
and an identical bound holds for $|\lan \phi| c_{\mb{k}\sigma}	[V,c^{\dg}_{\mb{k}\sigma}]|\phi\ran|$. 

To prove this version of Lemma 1, we first write $V = V_1 + V_2$ where
$V_1= \sum_{\mb{x},\sigma}(\ov{\mu}-\mu_{\mb{x}})n_{\mb{x}\sigma}$ contains the 
single-particle potential terms and $V_2 = \sum_{\mb{x}}u_{\mb{x}}n_{\mb{x}\up}n_{\mb{x}\down}$
contains the Hubbard interaction. By the triangle inequality we have
\beq
	|\lan \phi| c^{\dg}_{\mb{k}\sigma}	[V,c_{\mb{k}\sigma}]|\phi\ran| \leq
	|\lan \phi| c^{\dg}_{\mb{k}\sigma}	[V_1,c_{\mb{k}\sigma}]|\phi\ran| + |\lan \phi| c^{\dg}_{\mb{k}\sigma}	[V_2,c_{\mb{k}\sigma}]|\phi\ran|\ .
\eeq
We will now show that
that $|\lan \phi| c^{\dg}_{\mb{k}\sigma}	[V_1,c_{\mb{k}\sigma}]|\phi\ran| \leq s_{\mu}$ and
that $|\lan \phi| c^{\dg}_{\mb{k}\sigma}	[V_2,c_{\mb{k}\sigma}]|\phi\ran| \leq u_{\text{rms}}$.

To bound the term involving $V_2$ we follow almost the exact same steps as in the proof
of Lemma 1 in the main text. The only difference is that we now define $A^{(\al)}_{\mb{k}\mb{q}}$ by
\beq
	A^{(\al)}_{\mb{k}\mb{q}} := \sum_{\mb{x}}u_{\mb{x}}\lan\al|\td{n}_{\mb{x}\down}|\al\ran e^{-i(\mb{k}-\mb{q})\cdot\mb{x}}\ ,
\eeq
and we then find that
\begin{align}
	\sum_{\mb{q}}|A^{(\al)}_{\mb{k}\mb{q}}|^2 = |\Lambda|\sum_{\mb{x}}u_{\mb{x}}^2\lan\al|\td{n}_{\mb{x}\down}|\al\ran^2 \leq (u_{\text{rms}}|\Lambda|)^2\ ,
\end{align}
where the first equality again follows from the Plancherel theorem. We can now follow the same steps
as in the main text to conclude that 
$|\lan \phi| c^{\dg}_{\mb{k}\sigma}	[V_2,c_{\mb{k}\sigma}]|\phi\ran| \leq u_{\text{rms}}$.

Next, we prove the bound involving $V_1$. The proof in this case is much simpler and we do not
need to use a decomposition of the state $|\phi\ran$ like we had in 
Eq.~18 of the main text (that decomposition was only needed to handle the
Hubbard interaction term). In this case we simply have
\beq
	c^{\dg}_{\mb{k}\sigma}	[V_1, c_{\mb{k}\sigma}] = \frac{1}{|\Lambda|}\sum_{\mb{x},\mb{q}}(\mu_{\mb{x}}-\ov{\mu})c^{\dg}_{\mb{k}\sigma}c_{\mb{q}\sigma}e^{-i(\mb{k}-\mb{q})\cdot\mb{x}}\ ,
\eeq
and we can write
\beq
	\lan\phi| c^{\dg}_{\mb{k}\sigma}	[V_1, c_{\mb{k}\sigma}]|\phi\ran = \frac{1}{|\Lambda|}\sum_{\mb{q}}A_{\mb{k}\mb{q}}M_{\mb{k}\mb{q}}
\eeq
where now
\begin{subequations}
\begin{align}
	A_{\mb{k}\mb{q}} &:= \sum_{\mb{x}}(\mu_{\mb{x}}-\ov{\mu}) e^{-i(\mb{k}-\mb{q})\cdot\mb{x}} \\
	M_{\mb{k}\mb{q}} &:= \lan\phi|c^{\dg}_{\mb{k}\sigma}c_{\mb{q}\sigma}|\phi\ran\ .
\end{align}
\end{subequations}
Since $|\phi\ran$ is a normalized state, the same manipulations from the main text (Cauchy-Schwarz, 
Plancherel theorem, Fermi statistics) immediately lead to the desired bound
$|\lan \phi| c^{\dg}_{\mb{k}\sigma}	[V_1,c_{\mb{k}\sigma}]|\phi\ran| \leq s_{\mu}$.

This completes the proof of Theorem 1 and Lemma 1 for the family of generalized Hubbard models defined in 
Eq.~\eqref{eq:gen-Hubbard}.

\subsection{Ground state result}

We now prove our results for the momentum occupation numbers in the
ground state of the generalized Hubbard model \eqref{eq:gen-Hubbard}. In particular,
Theorem 2 from the main text will follow from a special case of the more general results that
we state and prove here. 

We start by
stating our ground state result. For this result recall that, even
if the ground state of this model is degenerate, we can always choose a basis for the space of ground
states such that each state is a simultaneous eigenstate of $\mathcal{N}_{\up}$ and 
$\mathcal{N}_{\down}$. In the statement of our result we assume that the
ground state $|\psi\ran$ is chosen
from such a basis in the case where the model has a ground state degeneracy. As in the main text, we 
denote the eigenvalue of $\mathcal{N}_{\sigma}$ for $|\psi\ran$ by $N_{\sigma}$, 
$\mathcal{N}_{\sigma}|\psi\ran = N_{\sigma}|\psi\ran$, and then 
$\rho_{\sigma} = N_{\sigma}/|\Lambda|$ is the filling fraction for spin $\sigma$.

\textit{Theorem 2 (generalized Hubbard, ground state):} In any ground state of the
Hamiltonian \eqref{eq:gen-Hubbard} the momentum occupation numbers $\lan n_{\mb{k}\sigma}\ran$ obey 
\begin{subequations}
\begin{align}
	\lan n_{\mb{k}\sigma}\ran &\leq \frac{s_{\mu} + |\ov{u}|\sqrt{\rho_{\ov{\sigma}}} + s_u}{\ep_{\mb{k}} - \ov{\mu} + \ov{u}\rho_{\ov{\sigma}}}\ ,\ \text{ if }\ \ep_{\mb{k}} - \ov{\mu} + \ov{u}\rho_{\ov{\sigma}} > 0 \\
	1 - \lan n_{\mb{k}\sigma}\ran &\leq \frac{s_{\mu} + |\ov{u}|\sqrt{\rho_{\ov{\sigma}}} + s_u}{|\ep_{\mb{k}} - \ov{\mu} + \ov{u}\rho_{\ov{\sigma}}|}\ ,\ \text{ if }\ \ep_{\mb{k}} - \ov{\mu} + \ov{u}\rho_{\ov{\sigma}} < 0\ ,
\end{align}
\end{subequations}
where $\ov{\sigma}$ is the opposite of $\sigma$ (e.g., $\ov{\up} = \down$).

The proof of this ground state result differs from the proof of our non-zero temperature results in a few
important ways, and we highlight the differences below. We only discuss the proof of the first 
inequality in Theorem 2 for the case of spin-up, as the proofs of the remaining cases are very similar.

The proof again starts with a local version of the variational principle, but in this case we only need
to use the familiar variational principle for the ground state. In particular, we use the fact that,
for any operator $\mathcal{O}$, the variational principle
implies that $\lan\mathcal{O}^{\dg}[H,\mathcal{O}]\ran \geq 0$, where
the expectation is taken in the ground state $|\psi\ran$. If we choose 
$\mathcal{O} = c_{\mb{k}\up}$, then we find that
\beq
	-(\ep_{\mb{k}}-\ov{\mu})\lan n_{\mb{k}\up}\ran + \lan c^{\dg}_{\mb{k}\up}[V,c_{\mb{k}\up}]\ran \geq 0 \ .
	\label{eq:variational-thm}
\eeq
The next step is to analyze the term $\lan c^{\dg}_{\mb{k}\up}[V,c_{\mb{k}\up}]\ran$. 
To do this we use a new
decomposition of $V$ as $V = V_1' + V_2' + V_3'$, where
$V_1' = V_1 = \sum_{\mb{x},\sigma}(\ov{\mu}-\mu_{\mb{x}})n_{\mb{x}\sigma}$ like before, but now
$V_2' = \ov{u}\sum_{\mb{x}}n_{\mb{x}\up}n_{\mb{x}\down}$ contains the average Hubbard interaction and
$V_3'= \sum_{\mb{x}}(u_{\mb{x}}-\ov{u})n_{\mb{x}\up}n_{\mb{x}\down}$ contains the deviation of the 
Hubbard interaction from its average.

We start with the term containing $V_2'$. After some algebra we find that
\begin{align}
	c^{\dg}_{\mb{k}\up}[V_2',c_{\mb{k}\up}] = -\frac{\ov{u}}{|\Lambda|}\sum_{\mb{x},\mb{q}}\td{c}^{\dg}_{\mb{k}\up}\td{c}_{\mb{q}\up}\td{n}_{\mb{x}\down}e^{-i(\mb{k}-\mb{q})\cdot\mb{x}}\ , 
	\label{eq:expectation-formula}
\end{align}
where the operators $\tilde{c}_{\mb{k}\sigma}$ (fermions with tildes) were defined in the main text.
Next, we split the expectation $\lan c^{\dg}_{\mb{k}\up}[V_2',c_{\mb{k}\up}]\ran$ into two
pieces by separating out the $\mb{q}=\mb{k}$ term in the sum. The contribution from the $\mb{q}=\mb{k}$
term is 
\beq
	-\frac{\ov{u}}{|\Lambda|}\lan\td{n}_{\mb{k}\up}\sum_{\mb{x}}\td{n}_{\mb{x}\down}\ran = -\frac{\ov{u}}{|\Lambda|}\lan n_{\mb{k}\up}\mathcal{N}_{\down}\ran = - \ov{u}\rho_{\down}\lan n_{\mb{k}\up}\ran\ .
\eeq
[Recall that any number operator with a tilde is equal to the corresponding operator without the tilde.]
If we also define the quantity 
$\lan c^{\dg}_{\mb{k}\up}[V_2',c_{\mb{k}\up}]\ran_{\text{not }\mb{k}}$ by
\beq
\lan c^{\dg}_{\mb{k}\up}[V_2',c_{\mb{k}\up}]\ran_{\text{not }\mb{k}} := -\frac{\ov{u}}{|\Lambda|}\sum_{\mb{x},\mb{q}\neq \mb{k}}\td{c}^{\dg}_{\mb{k}\up}\td{c}_{\mb{q}\up}\td{n}_{\mb{x}\down}e^{-i(\mb{k}-\mb{q})\cdot\mb{x}}\ ,
\eeq
then we can now write Eq.~\eqref{eq:variational-thm} in the form
\beq
	(\ep_{\mb{k}}-\ov{\mu} + \ov{u}\rho_{\down})\lan n_{\mb{k}\up}\ran \leq \lan c^{\dg}_{\mb{k}\up}[V_1',c_{\mb{k}\up}]\ran + \lan c^{\dg}_{\mb{k}\up}[V_2',c_{\mb{k}\up}]\ran_{\text{not }\mb{k}} + \lan c^{\dg}_{\mb{k}\up}[V_3',c_{\mb{k}\up}]\ran
	\ . \label{eq:refined-ineq}
\eeq 
This inequality of course implies that 
\beq
	(\ep_{\mb{k}}-\ov{\mu} + \ov{u}\rho_{\down})\lan n_{\mb{k}\up}\ran \leq |\lan c^{\dg}_{\mb{k}\up}[V_1',c_{\mb{k}\up}]\ran| + |\lan c^{\dg}_{\mb{k}\up}[V_2',c_{\mb{k}\up}]\ran_{\text{not }\mb{k}}| + |\lan c^{\dg}_{\mb{k}\up}[V_3',c_{\mb{k}\up}]\ran|\ ,
\eeq
which will be useful when $\ep_{\mb{k}}-\ov{\mu} + \ov{u}\rho_{\down}>0$.
We can now use the same techniques that we used to prove Lemma 1 to show that 
$|\lan c^{\dg}_{\mb{k}\up}[V_1',c_{\mb{k}\up}]\ran| \leq s_{\mu}$ and that
$|\lan c^{\dg}_{\mb{k}\up}[V_3',c_{\mb{k}\up}]\ran| \leq s_{u}$.

The last step of the proof is to show that 
$|\lan c^{\dg}_{\mb{k}\up}[V_2',c_{\mb{k}\up}]\ran_{\text{not }\mb{k}}| \leq |\ov{u}|\sqrt{\rho_{\down}}$. 
To prove this bound we follow most of the proof of Lemma 1 from the main text, but we then follow a
different procedure at the end. Specifically, we start by decomposing the ground state
$|\psi\ran$ as
\beq
	|\psi\ran = \sum_{a,\al}W_{a\al}|a\ran \otimes |\alpha\ran\ ,
\eeq
where now the states $|a\ran$ are a real space occupation number basis for the $N_{\up}$-particle sector of
$\tilde{\mathcal{F}}_{\up}$, and the states $|\al\ran$ are a real space occupation number basis for the 
$N_{\down}$-particle sector of $\tilde{\mathcal{F}}_{\down}$ (the main difference from the main text
is that we are now working in a sector with a fixed number of particles of each spin). 
If we now follow the same steps as in the main text, then we will arrive at the inequality
\beq
	|\lan c^{\dg}_{\mb{k}\up}[V_2',c_{\mb{k}\up}]\ran_{\text{not }\mb{k}}| \leq \frac{1}{|\Lambda|}\sum_{\al}\sqrt{\left(\sum_{\mb{q}\neq\mb{k}}|A^{(\al)}_{\mb{k}\mb{q}}|^2\right)\left(\sum_{\mb{q}\neq\mb{k}}|M^{(\al)}_{\mb{k}\mb{q}}|^2\right)}\ , 
\eeq
where now
\begin{subequations}
\begin{align}
	A^{(\al)}_{\mb{k}\mb{q}} &:= \ov{u}\sum_{\mb{x}}\lan\al|\td{n}_{\mb{x}\down}|\al\ran e^{-i(\mb{k}-\mb{q})\cdot\mb{x}} \\
	M^{(\al)}_{\mb{k}\mb{q}} &:= \lan\chi_{\al}|\td{c}^{\dg}_{\mb{k}\up}\td{c}_{\mb{q}\up}|\chi_{\al}\ran\ ,
\end{align}
\end{subequations}
and we again have $|\chi_{\al}\ran := \sum_a W_{a\al}|a\ran$ and 
$\sum_{\al}\lan\chi_{\al}|\chi_{\al}\ran = 1$. 

Next, we again need to bound the sums involving $A^{(\al)}_{\mb{k}\mb{q}}$ and $M^{(\al)}_{\mb{k}\mb{q}}$.
To do this we first note that 
$\sum_{\mb{q}\neq\mb{k}}|A^{(\al)}_{\mb{k}\mb{q}}|^2\leq\sum_{\mb{q}}|A^{(\al)}_{\mb{k}\mb{q}}|^2$ 
(i.e., we add back in the $\mb{q}=\mb{k}$ term), and
likewise for the sum involving $M^{(\al)}_{\mb{k}\mb{q}}$. Next, we bound 
$\sum_{\mb{q}}|M^{(\al)}_{\mb{k}\mb{q}}|^2$ using Fermi statistics exactly as in the main text. Finally, 
for the sum over $A^{(\al)}_{\mb{k}\mb{q}}$ we first have
\beq
	\sum_{\mb{q}}|A^{(\al)}_{\mb{k}\mb{q}}|^2 = \ov{u}^2|\Lambda|\sum_{\mb{x}}\lan\al|\td{n}_{\mb{x}\down}|\al\ran^2\ .
\eeq
Then, since $\lan\al|\td{n}_{\mb{x}\down}|\al\ran^2 = 
\lan\al|\td{n}_{\mb{x}\down}|\al\ran$ (because $\lan\al|\td{n}_{\mb{x}\down}|\al\ran$ equals $0$ or $1$), 
we have
\beq
	\sum_{\mb{q}}|A^{(\al)}_{\mb{k}\mb{q}}|^2 = \ov{u}^2 |\Lambda|\sum_{\mb{x}}\lan\al|\td{n}_{\mb{x}\down}|\al\ran = \ov{u}^2 |\Lambda| N_{\down}\ .
\eeq
Putting these results together then yields the bound 
$|\lan c^{\dg}_{\mb{k}\up}[V_2',c_{\mb{k}\up}]\ran_{\text{not }\mb{k}}| \leq |\ov{u}|\sqrt{\rho_{\down}}$.

After all of this work, we end up with the inequality 
\beq
	(\ep_{\mb{k}}-\ov{\mu} + \ov{u}\rho_{\down})\lan n_{\mb{k}\up}\ran \leq s_{\mu} + |\ov{u}|\sqrt{\rho_{\down}} + s_u\ ,
\eeq
and then the first inequality in Theorem 2 follows for any $\mb{k}$ that satisfies
$\ep_{\mb{k}}-\ov{\mu} + \ov{u}\rho_{\down} > 0$.

Finally, to derive the second inequality in Theorem 2 one should repeat this proof but choose 
$\mathcal{O} = c_{\mb{k}\up}^{\dg}$ instead of $\mathcal{O} = c_{\mb{k}\up}$ in the first step where
we used the variational principle.

\section{Results for generalized Kondo models}

In this section we present the proofs of our results for the Kondo model, namely the proofs of Theorem 1 (the 
Kondo part) and Theorem 3 from the main text. We will actually prove these results for a more general
Kondo model that includes more than one impurity spin. 

We now consider generalized Kondo models that feature a finite number $M$ of impurity spins 
$\vec{S}_1,\dots,\vec{S}_M$ at arbitrary locations and with arbitrary values of their spin and Kondo coupling. 
The $i$th impurity spin, for $i \in\{1,\dots,M\}$, has spin $s_i \in\{1/2,1,3/2,\dots\}$. These
spins couple to the fermions on $M$ distinct lattice sites $\mb{x}_1,\dots,\mb{x}_M$. 
Finally, the spin $\vec{S}_i$ couples to the fermion at site $\mb{x}_i$ with a Kondo coupling $J_i$. The
Hamiltonian for this generalized Kondo model takes the form $H_{\text{Kondo}} = H_0 + V$, where
$H_0 = \sum_{\mb{k},\sigma}(\ep_{\mb{k}}-\mu)n_{\mb{k}\sigma}$ and
\begin{subequations}
\label{eq:Kondo-V}
\beqa
	V &=& \sum_{i=1}^M V_i \\
	V_i &=&  J_i \vec{S}_i\cdot \left(\sum_{\tau,\tau'} c^{\dg}_{\mb{x}_i\tau}\frac{\vec{\sigma}_{\tau\tau'}}{2}c_{\mb{x}_i\tau'}\right)\ . 
\eeqa
\end{subequations}
Our main results for this generalized Kondo model are as follows.

\textit{Theorem 1 (generalized Kondo, non-zero temperature):} Let $f_{\mb{k}}$ denote the Fermi-Dirac 
distribution with chemical potential $\mu$, 
$f_{\mb{k}} = (e^{\beta (\ep_{\mb{k}} - \mu)} + 1)^{-1}$.
For any $\beta < \infty$, the momentum occupation numbers $\lan n_{\mb{k}\sigma}\ran$ for 
$H_{\text{Kondo}}$ obey
\beq
	-\delta f_{\mb{k}} \leq \lan n_{\mb{k}\sigma}\ran - f_{\mb{k}} \leq \delta(1 - f_{\mb{k}})\ ,
\eeq
where the constant $\delta$ is given by
\beq
	\delta = \beta \frac{3}{2\sqrt{|\Lambda|}}\sum_i |J_i|s_i\ .
\eeq

\textit{Theorem 3 (generalized Kondo, ground state):} In any ground state of $H_{\text{Kondo}}$ 
the momentum occupation numbers $\lan n_{\mb{k}\sigma}\ran$ obey 
\begin{subequations}
\begin{align}
	\lan n_{\mb{k}\sigma}\ran &\leq \frac{3}{2\sqrt{|\Lambda|}}\frac{\sum_i |J_i|s_i}{\ep_{\mb{k}} - \mu}\ \ ,\ \text{ if }\ \ep_{\mb{k}} - \mu  > 0 \\
	1 - \lan n_{\mb{k}\sigma}\ran &\leq \frac{3}{2\sqrt{|\Lambda|}}\frac{\sum_i |J_i|s_i}{|\ep_{\mb{k}} - \mu|}\ \ ,\ \text{ if }\ \ep_{\mb{k}} - \mu < 0\ .
\end{align}
\end{subequations}

The main structure of the proofs of these results is identical to the Hubbard model case. The only difference 
is in the specific bound on the terms involving the interaction, and so in this section we will only derive 
that specific bound. In particular, we will prove the following result.

\textit{Lemma 1 (generalized Kondo model):} 
Let $|\phi\ran$ be any normalized state, and let $V$ be the Kondo interaction term from
Eq.~\eqref{eq:Kondo-V}. Then for any $\mb{k}$ and $\sigma$ the expectation value
$\lan \phi| c^{\dg}_{\mb{k}\sigma}	[V,c_{\mb{k}\sigma}]|\phi\ran$ obeys the bound
\beq
	|\lan \phi| c^{\dg}_{\mb{k}\sigma}	[V,c_{\mb{k}\sigma}]|\phi\ran| \leq \frac{3}{2\sqrt{|\Lambda|}}\sum_i |J_i|s_i\ ,
\eeq
and an identical bound holds for $|\lan \phi| c_{\mb{k}\sigma}	[V,c^{\dg}_{\mb{k}\sigma}]|\phi\ran|$. 

We now prove this result for the case of spin-up. We start by writing out the Kondo interaction in 
more detail. For any index $i$ we have
\beq
	V_i = \frac{J_i}{2}\left(S^x_i c^{\dg}_{\mb{x}_i\up}c_{\mb{x}_i\down} - iS^y_i 
	c^{\dg}_{\mb{x}_i\up}c_{\mb{x}_i\down} + \text{H.c.} \right) + \frac{J_i}{2}S^z_i(n_{\mb{x}_i\up} - 
	n_{\mb{x}_i\down}) \ ,
\eeq
where H.c. = Hermitian conjugate. By the triangle inequality we have
$|\lan \phi| c^{\dg}_{\mb{k}\sigma}	[V,c_{\mb{k}\sigma}]|\phi\ran| \leq \sum_i |\lan \phi| 
c^{\dg}_{\mb{k}\sigma}	[V_i,c_{\mb{k}\sigma}]|\phi\ran|$, and the individual terms 
$c^{\dg}_{\mb{k}\up}[V_i,c_{\mb{k}\up}]$ take the form
\beq
	c^{\dg}_{\mb{k}\up}[V_i,c_{\mb{k}\up}] = \frac{J_i}{2\sqrt{|\Lambda|}}e^{-i\mb{k}\cdot\mb{x}_i}\left(-S^x_i c^{\dg}_{\mb{k}\up}c_{\mb{x}_i\down} + i S^y_i c^{\dg}_{\mb{k}\up}c_{\mb{x}_i\down} - S^z_i c^{\dg}_{\mb{k}\up}c_{\mb{x}_i\up}\right)\ .
\eeq
The triangle inequality then gives
\beq
	|\lan \phi| c^{\dg}_{\mb{k}\up}[V_i,c_{\mb{k}\up}]|\phi\ran| \leq \frac{|J_i|}{2\sqrt{|\Lambda|}}\left(|\lan \phi| S^x_i c^{\dg}_{\mb{k}\up}c_{\mb{x}_i\down}|\phi\ran| + |\lan\phi| S^y_i c^{\dg}_{\mb{k}\up}c_{\mb{x}_i\down}|\phi\ran| + |\lan\phi| S^z_i c^{\dg}_{\mb{k}\up}c_{\mb{x}_i\up}|\phi\ran|\right)\ .
\eeq

To proceed further we now derive a bound on the absolute value of a general expectation value of the form
$\lan\phi| S^{\al}_i c^{\dg}_{\mb{k}\sigma}c_{\mb{x}_i\tau}|\phi\ran$ where $\al \in \{x,y,z\}$ and
$\sigma,\tau\in\{\up,\down\}$. To start, the Cauchy-Schwarz inequality gives 
\beq
	|\lan\phi| S^{\al}_i c^{\dg}_{\mb{k}\sigma}c_{\mb{x}_i\tau}|\phi\ran| \leq \sqrt{\lan\phi|(S^{\al}_i)^2|\phi\ran\lan\phi|c^{\dg}_{\mb{x}_i\tau} c_{\mb{k}\sigma} c^{\dg}_{\mb{k}\sigma}c_{\mb{x}_i\tau}|\phi\ran}\ .
\eeq
Since we are working with an impurity of spin $s_i$, we have 
$\lan\phi|(S^{\al}_i)^2|\phi\ran \leq s_i^2$. For the other term we have
\beqa
	\lan\phi|c^{\dg}_{\mb{x}_i\tau} c_{\mb{k}\sigma} c^{\dg}_{\mb{k}\sigma}c_{\mb{x}_i\tau}|\phi\ran &=& \lan\phi|c^{\dg}_{\mb{x}_i\tau} (1-n_{\mb{k}\sigma})c_{\mb{x}_i\tau}|\phi\ran \nnb \\
	&\leq& \lan\phi|c^{\dg}_{\mb{x}_i\tau} c_{\mb{x}_i\tau}|\phi\ran \nnb \\
	&\leq& 1\ ,
\eeqa
where we used the fact that the operators $1-n_{\mb{k}\sigma}$ and 
$c^{\dg}_{\mb{x}_i\tau} c_{\mb{x}_i\tau} = n_{\mb{x}_i\tau}$ both have a maximum eigenvalue equal to $1$ 
(this is where the Fermi statistics of the particles is used to derive our result for the Kondo model).
Putting these results together yields the bound
\beq
	|\lan\phi| S^{\al}_i c^{\dg}_{\mb{k}\sigma}c_{\mb{x}_i\tau}|\phi\ran| \leq s_i\ ,
\eeq
which is enough to complete the proof of the stated bound on 
$|\lan \phi| c^{\dg}_{\mb{k}\sigma}[V,c_{\mb{k}\sigma}]|\phi\ran|$.

\section{Eigenvalues of fermion density matrices}

In this last section we present a short proof of the fact about fermion density matrices that we stated
in the main text and used in the proof of Lemma 1.

We first recall the setup that we had in the main text. We considered a set of 
fermion creation and annihilation operators $c_i$, $c^{\dg}_i$ obeying $\{c_i,c_j\}=0$ and 
$\{c_i,c^{\dg}_j\} = \delta_{ij}$, where the indices $i$ and $j$
take values in some finite index set $\mathcal{I}$. We then picked a particular state
$|\chi\ran$ in the Fock space
of these fermions, and we used it to define a Hermitian matrix $M$ whose matrix elements are given by
$M_{ij} := \lan\chi| c^{\dg}_i c_j|\chi\ran$. In the main text we claimed
that, if $\lambda$ is any eigenvalue of $M$, then $0\leq \lambda\leq \lan\chi|\chi\ran$ (we do not
assume that $|\chi\ran$ is normalized). We now present a short proof of this result.

Let $v_i$ be the components of a normalized eigenvector of $M$ with eigenvalue $\lambda$, 
so $\sum_j M_{ij}v_j = \lambda v_i$ for all $i$ and $\sum_i |v_i|^2 = 1$. If we define a new
fermion operator
$\mathcal{C}_v$ by $\mathcal{C}_v := \sum_i v_i c_i$, then we have  
$\lambda = \sum_{i,j}v^*_i M_{ij}v_j = \lan\chi|\mathcal{C}^{\dg}_v \mathcal{C}_v|\chi\ran$ 
(the star denotes complex conjugation).
One can check that $\mathcal{C}_v$ obeys the standard anticommutation relations 
$\{\mathcal{C}_v, \mathcal{C}_v\} = 0$ and $\{\mathcal{C}_v, \mathcal{C}^{\dg}_v\} = 1$. Then the number
operator $\mathcal{C}^{\dg}_v \mathcal{C}_v$ has eigenvalues $0$ and $1$, which implies that
$0\leq \lambda\leq \lan\chi|\chi\ran$.